\def\year{2018}\relax
\documentclass[letterpaper]{article} 
\usepackage{aaai18}  
\usepackage{times}  
\usepackage{helvet}  
\usepackage{courier}  
\usepackage{url}  
\usepackage{graphicx}  
\usepackage{blindtext}
\usepackage{mathtools}
\usepackage{booktabs} 
\usepackage{amsfonts}
\usepackage{float}
\usepackage{flushend}
\begin{document}
%
\title{Latent Structure in Collaboration: the Case of Reddit r/place}

\author{J{\'e}r{\'e}mie Rappaz \\EPFL \\ jeremie.rappaz@epfl.ch \And Michele Catasta \\ Stanford University \\ pirroh@cs.stanford.edu \And Robert West \\ EPFL \\ robert.west@epfl.ch \And Karl Aberer\\ EPFL \\ karl.aberer@epfl.ch\\ 
}

\newcommand{\xhdr}[1]{\vspace{2mm}\noindent{{\bf #1}}}

\maketitle
\begin{abstract}
Many Web platforms rely on user collaboration to generate high-quality content: Wiki, Q\&A communities, etc. Understanding and modeling the different collaborative behaviors is therefore critical. However, collaboration patterns are difficult to capture when the relationships between users are not directly observable, since they need to be inferred from the user actions. In this work, we propose a solution to this problem by adopting a systemic view of collaboration. Rather than modeling the users as independent actors in the system, we capture their coordinated actions with embedding methods which can, in turn, identify shared objectives and predict future user actions.

To validate our approach, we perform a study on a dataset comprising more than 16M user actions, recorded on the online collaborative sandbox Reddit~\textit{r/place}. Participants had access to a drawing canvas where they could change the color of one pixel at every fixed time interval. Users were not grouped in teams nor were given any specific goals, yet they organized themselves into a cohesive social fabric and collaborated to the creation of a multitude of artworks. Our contribution in this paper is multi-fold: i) we perform an in-depth analysis of the Reddit~\textit{r/place} collaborative sandbox, extracting insights about its evolution over time; ii) we propose a predictive method that captures the latent structure of the emergent collaborative efforts; and iii) we show that our method provides an interpretable representation of the social structure.
\end{abstract}

\noindent 

\section{Introduction}

Human beings left free to act according to their own will seem to often produce spontaneous order. This phenomenon has been observed in various contexts, ranging from city traffic flow to self-organizing economy. Those examples suggest that, even with different goals, a multitude of individuals interacting with one another often tend to avoid disorder, letting some underlying structure emerge. The Web is no exception: many Web initiatives rely on this principle, such as Question Answering platforms, collaborative code platforms, or even larger platforms such as Wikipedia. Characterizing the way people interact and organize themselves is a necessary step to understand users' behavior when little or no rules are enforced. Broadening our understanding of collaboration and self-organization can, therefore, have a practical impact on the design of applications that are built for large populations of users, which underlines the importance of expanding our ensemble of methods to study such phenomena.

\begin{figure}
	\centering
    \includegraphics[width=1.0\linewidth]{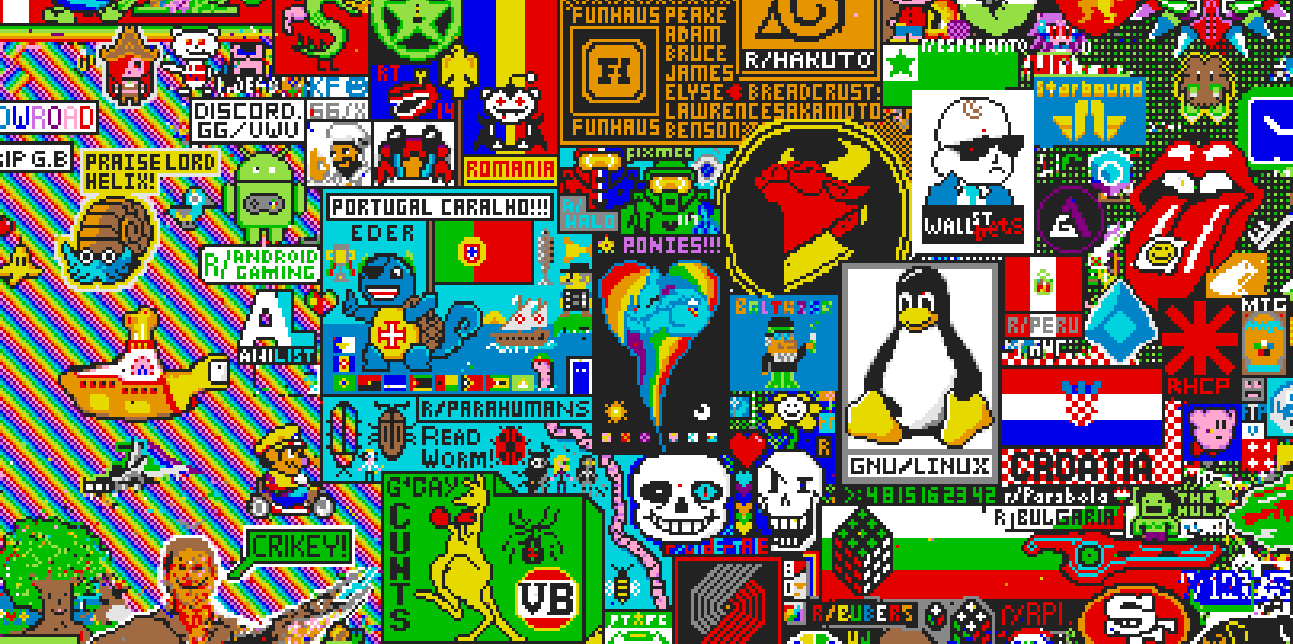}
    \label{fig:final_canvas}
    \caption{A portion of the final canvas produced by the contributors of Reddit r/place}
\end{figure}

The study of collaboration is, however, being hampered by the inherent complexity of the task: the phenomenon is hard to be analyzed in realistic scenarios as it requires to capture complex interactions between actors. Therefore, the dynamics of such efforts have been mostly studied in scenarios where the relationships among users are directly observable. The problem of community detection in collaboration networks, for example, relies on an existing graph to segment it into subgroups of users. However, in practical scenarios, the relationships between users are often not explicit. In such cases, the analysis can rely on human judgment \cite{Majchrzak2005PerceivedIC}, however with severe limitations in terms of size of the environment that can be analyzed, both in terms of number of users and interactions. Therefore, we propose in this work a data-driven method to infer the user relationships from their interactions in the environment. 

\textbf{In this work} we propose a methodology to analyze user activities in a simple virtual environment, with the goal of characterizing collaboration patterns that emerged from it. In a second step, we propose a predictive method that captures the latent structure of user cooperation, and evaluate our approach on its capacity to predict future user actions. Our study focuses on a virtual sandbox in which users receive no particular directives, and are given the freedom to act as they want. In particular, we propose to investigate the behavioral patterns observed in Reddit\footnote{https://www.reddit.com} \textit{r/place}\footnote{https://www.reddit.com/r/place/}, an online canvas where users were allowed to change the color of only one pixel at every fixed time interval (during a total of 72 hours). Being part of Reddit, a discussion platform with more than 230M unique users per month, this collaborative project received a massive engagement of over a million users (see Figure~\ref{fig:final_canvas} for a glimpse of the final canvas). 

In order to establish a predictive model of user behavior, we consider the sandbox as a complex social system, i.e., a system inherently difficult to model due to the large amount of interdependencies between its parts. Previous research in the field of Complexity Science \cite{bar2002general} hypothesized that the nature of such systems is favorable to the emergence of global behaviors, arising from the local interactions of the actors. Following this evidence, we propose a model that assesses the likelihood of a user interaction by observing its social context. In other terms, we propose a predictive model that captures inter-user relationships instead of modeling independent behaviors.





The rest of this paper is organized as follow. In section~\ref{sec:related}, we describe relevant pieces of work. In section~\ref{sec:data}, we describe the dataset. In section~\ref{sec:analysis}, we analyze the evolution of the dataset over time. In section~\ref{sec:model}, we propose a predictive model and we evaluate its performance in section~\ref{sec:results}. In section~\ref{sec:segmentation}, we propose a way to visualize the different collaborative efforts.

\begin{figure*}
\hfill%
\begin{minipage}[t]{0.27\linewidth}
    \includegraphics[width=\linewidth]{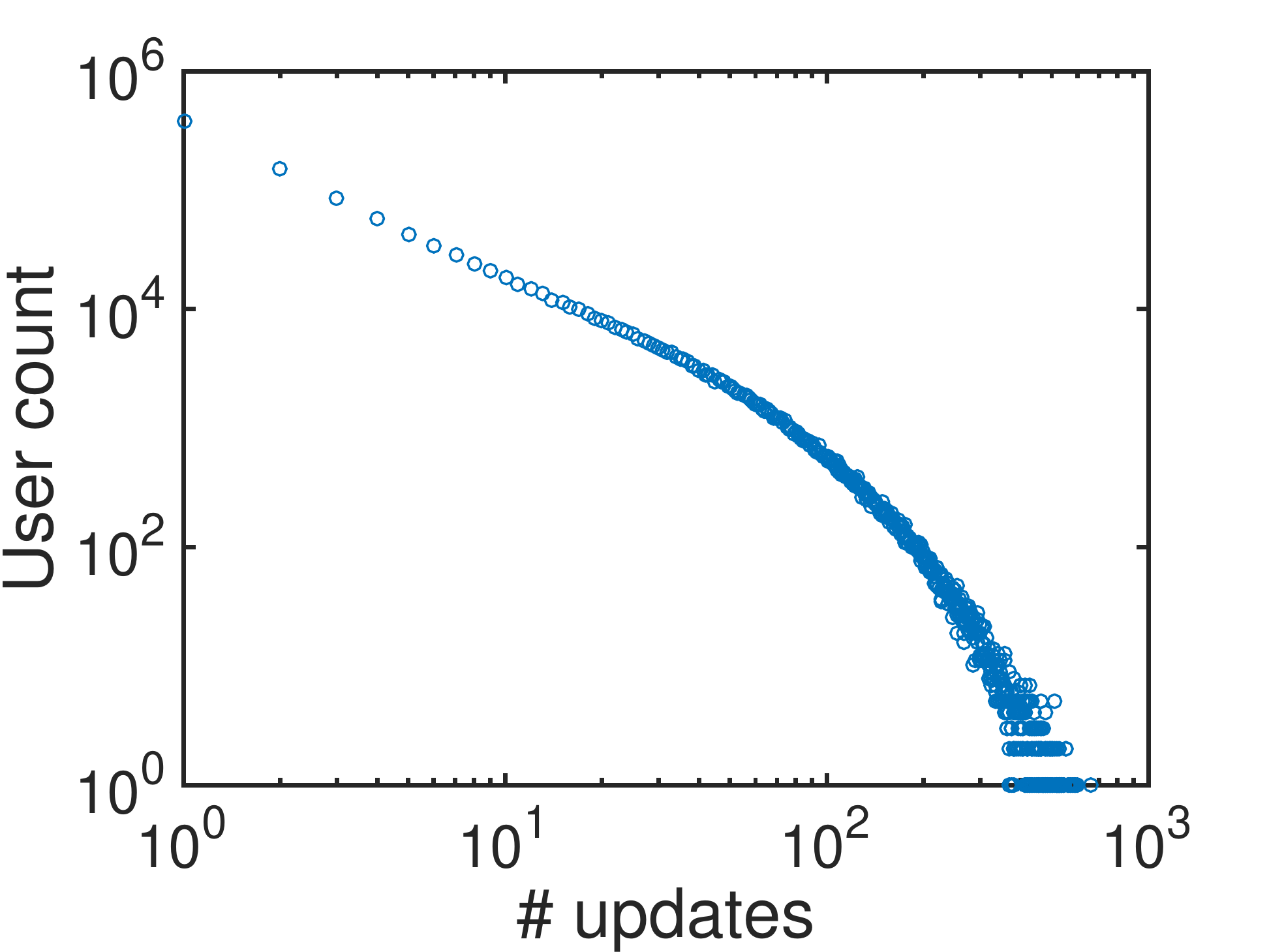}
\end{minipage}%
\hfill%
\begin{minipage}[t]{0.27\linewidth}
    \includegraphics[width=\linewidth]{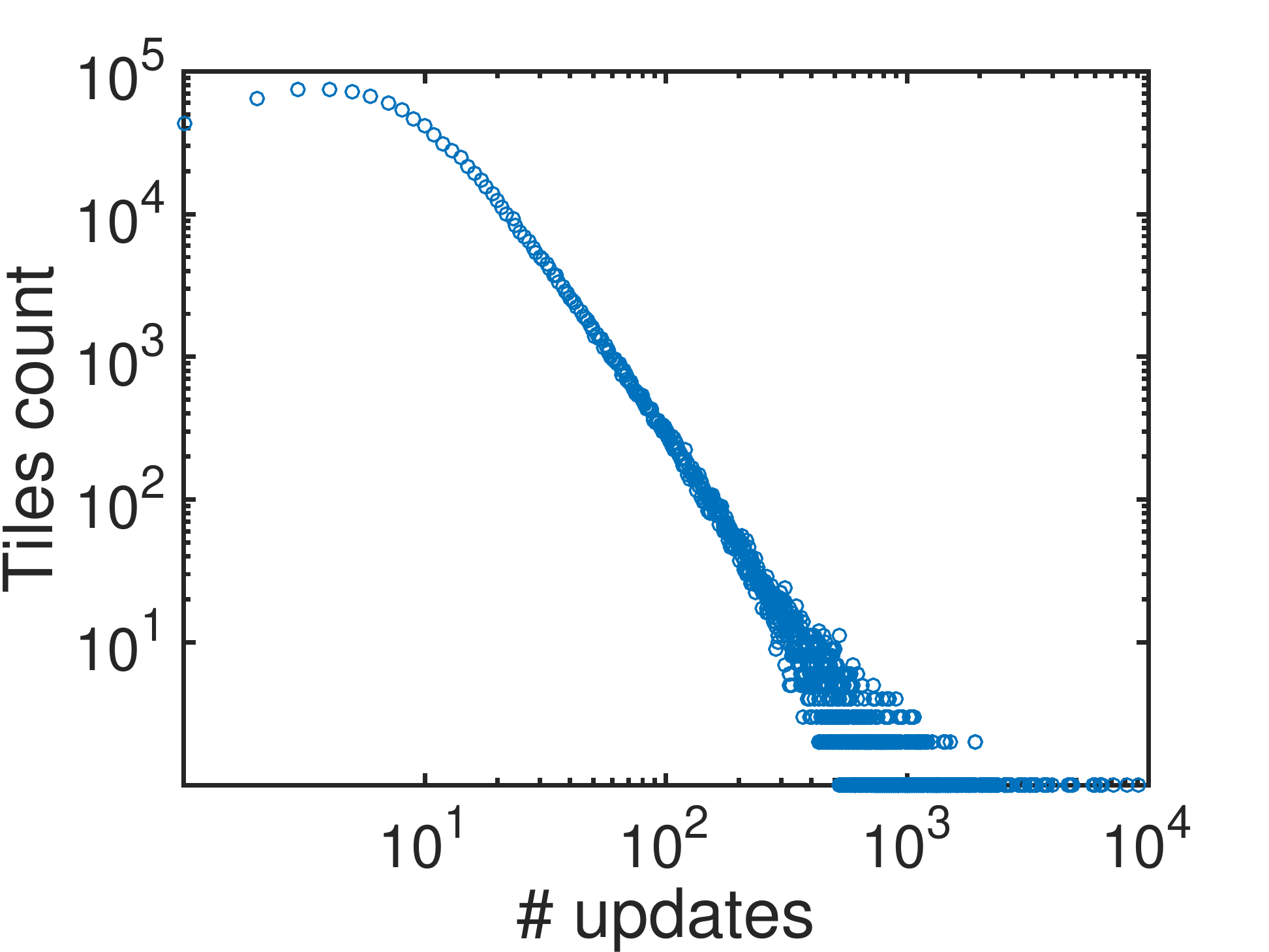}
\end{minipage}
\hfill%
\begin{minipage}[t]{0.32\linewidth}
    \includegraphics[width=\linewidth]{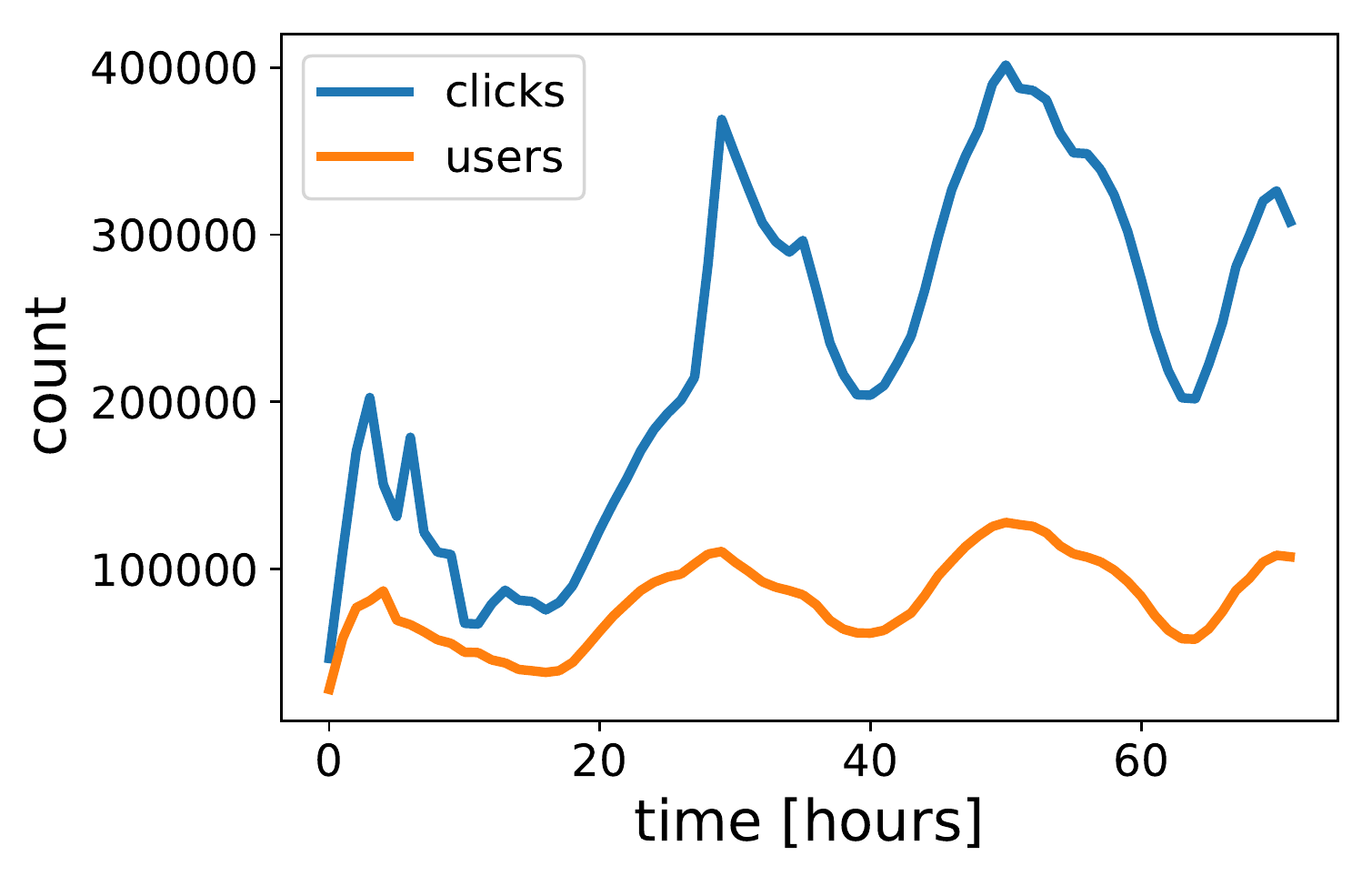}
\end{minipage}
\hfill%
\caption{Distribution of the number of updates for each user (left). Distribution of the number of updates per tile (middle). Evolution of the number of clicks and the number of unique users, computed per hour (right)}
\label{fig:distribs}
\end{figure*}

\section{Related Work}
\label{sec:related}

Networks of collaboration between scientists have been studied by Newman \cite{Newman2001ScientificCN}. The authors argue that simple unweighted networks are unable to capture the strength of collaboration ties and propose a method to model the strength of collaboration by relying on the number of co-authored papers. The same author has later studied various properties of such networks \cite{Newman2001ScientificCN}. Ramasco et al. \cite{ramasco2004self} have studied collaboration networks from an evolving and self-organizing perspective. Behavioral experiments on the ability to solve problems collaboratively have been conducted by Kearns \cite{kearns2012experiments}. Online collaboration with different network topologies has been studied by Suri and Watts \cite{suri2011cooperation}. The exploration-exploitation trade-off in a collaborative problem solving task has been discussed by Mason and Watts \cite{mason2012collaborative}. Kittur and Kraut \cite{kittur2008harnessing} studied various types of collaboration taking place between Wikipedia editors and measured the impact on quality of the resulting articles.

The study, as well as the interpretation of proximity data from a social perspective has been a prolific research area. Recent studies 
\cite{Sekara2014TheSO}, have extracted social network properties from proximity sensor data. In particular, the authors propose a method to distinguish between strong and weak social ties, using the Bluetooth signal strength of users' cellphones. The authors observe that weak links, i.e. the interactions that have been observed less than once per day, have a lower probability of being observed at later times. Collaboration patterns between university students, collaborating in teams for their course assignments, have also been studied \cite{Montjoye2014TheSO}. The authors consider the time spent in physical proximity, using university wifi logs, as a proxy to measure ties between students. Their analysis suggests that only strong ties matter in order to predict team performances. We also notice that the study of social properties from positional tracking is not limited to the human species, as a colony of ants as been recently tracked, at individual level, revealing complex hierarchical social structures \cite{Mersch1090}.

The phenomenon of emergence has been studied in different domains, notably in the field of Complexity Science and in the context of agent based modeling. The term emergence has various definition across fields \cite{doi:10.1162/106454603321489518}, alike complexity \cite{Gershenson2012ComplexityAI} from which emergence has been suggested to arise from. Emergence generally refers to system-wide behaviors that cannot be explained by the sum of individual behaviors. Moreover, means of modeling emergence are still subject to debate. Counting interaction between agents is, however, a widely used method to infer complex behaviors in a system \cite{mataric1993designing}. 

The analysis of virtual behavior has been suggested as being a valid proxy for the study of real-life behaviors. High-level social behaviors, such as the bystander effect, have been observed inside a simple video-game based virtual environment \cite{Kozlov2010RealBI}. Social interactions in Massively Multilayer Online games have been studied by Cole et al. \cite{cole2007social}. They suggest such games to be favorable for teamwork.

The task of community detection has been a well-studied problem, whose goal is to assign users to communities \cite{Rosvall2008MapsOR}. The most relevant line of research is probably the task of detecting \textit{overlapping} communities, whose members can be part of multiple groups. Those lines of research have made use of Matrix Factorization methods in order to relax the assumption of communities being disjoint \cite{Zhang2012OverlappingCD} \cite{Yang2013OverlappingCD}. Methods providing a direct way to embed the nodes of a network, thus generalizing the notion of network neighborhood, have recently been proposed \cite{grover2016node2vec}.

The modeling of personalized decision processes has many application, notably in Collaborative Filtering, which generally assumes that future user actions can be predicted by collecting historical data from many users. Such applications have made extensive use of Matrix Factorization methods that have been popularized during the Netflix Prize \cite{Bell2007TheBS}. Traditional techniques were mostly relying on ratings data but recent advances of the field have focused their attention on the problem of One-Class Collaborative Filtering \cite{Pan2008OneClassCF}, that is the task of learning from positive interactions only. More recently, Rendle et al. \cite{Rendle2009BPRBP} proposed a variation of the method that is particularly suited for the modeling of one-class datasets. A link between recommendation and collaboration has been explored by Tang et al. \cite{Tang2012CrossdomainCR}. In particular, the authors tackled the problem of scientific collaboration recommendation. 

\xhdr{Research Questions:}
 Given the work above, several research questions have remained unanswered:
\begin{description}
\item \textbf{RQ1:} Are local coordinated user activities predictive of future user actions? 
\item \textbf{RQ2:} Is the latent representation of our model interpretable?
\item  \textbf{RQ3:} Is the product of users' collaboration segmentable relying on implicit social signal exclusively?
\end{description}

\section{Data}
\label{sec:data}
In April 2017, the discussion platform \textit{Reddit} launched \textit{Place}, an online canvas of 1000-by-1000 pixels, designed as a social experiment. Reddit users were allowed to change the color of one pixel at every fixed time interval (the interval varied from 5 to 20 minutes during the events). The event lasted 72 hours and received a massive engagement from more than 1.2M unique users. Users collaborated to create various artworks by either directly interacting with the canvas or by coordinating their actions from the discussion platform.

The full dataset of events has been made publicly available. It contains more than 16M events and includes, for each event, the position of the click in the canvas, the chosen color and a unique user identifier. Note that usernames have been anonymized using a hash function, making impossible to link users to their respective \textit{Reddit} profile.

After the end of the event, Reddit users launched an initiative to segment and detail the various artworks of the final canvas. Their crowd-sourced effort resulted in an atlas publicly available online \footnote{https://draemm.li/various/place-atlas/}. In total, 1493 artworks have been identified by the community. In their original format, those annotations contained overlapping regions, coming from noisy segmentations or from users annotating subparts of the artworks. In this work, we do not consider the hierarchy of the artworks and, therefore, we have to further process the set of artworks through manual annotation with the following strategy: if two artworks have a non-zero intersection, we give priority to the smallest one and remove the overlapping region from the largest one. Then, we manually and iteratively select coherent artworks made of a single region. The result is a set of 830 non-overlapping artworks covering $84.2\%$ of the canvas.

\section{Analysis}
\label{sec:analysis}
In this section, we propose a general analysis of the event by shining a light on user behaviors within the online canvas.

We first observe, in figure~\ref{fig:distribs} (left), the activity distribution of the users. This distribution highlight the presence of few power-users and a vast majority of users performing a moderate number of clicks. In figure~\ref{fig:distribs} (middle), we observe the same type of distribution for the number of updates performed on every pixel. As few pixels have been highly disputed, the large majority of them have only been updated a few times. For example, 4.35\% of the pixels have been updated only once during the entire event. The fact that the number of updates is not uniform over the set of pixels suggests that there exists a latent structure in the users' decision process.

\begin{figure}[H]
\begin{minipage}[t]{0.48\linewidth}
    \includegraphics[width=\linewidth]{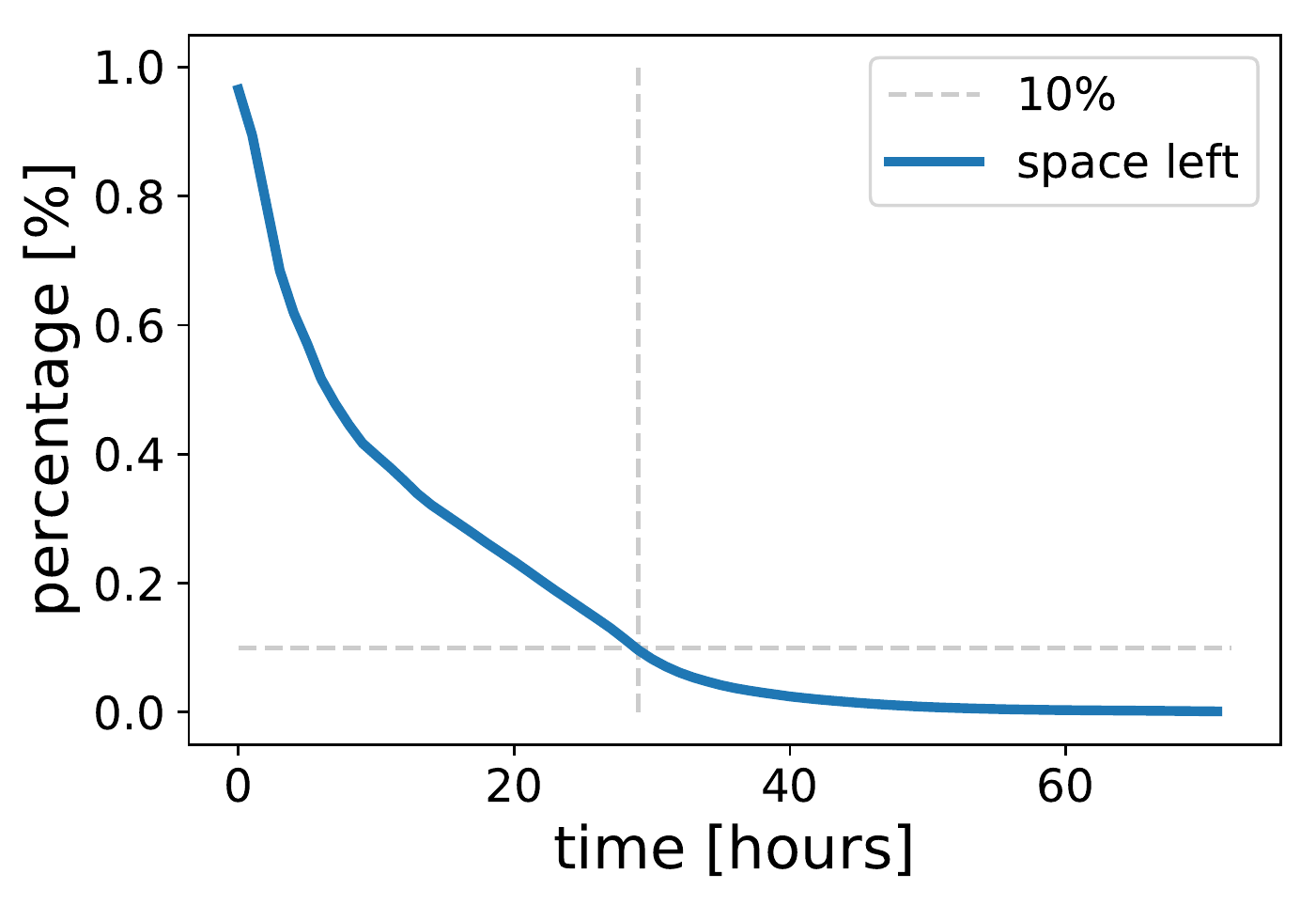}
\end{minipage}%
\hfill%
\begin{minipage}[t]{0.48\linewidth}
    \includegraphics[width=\linewidth]{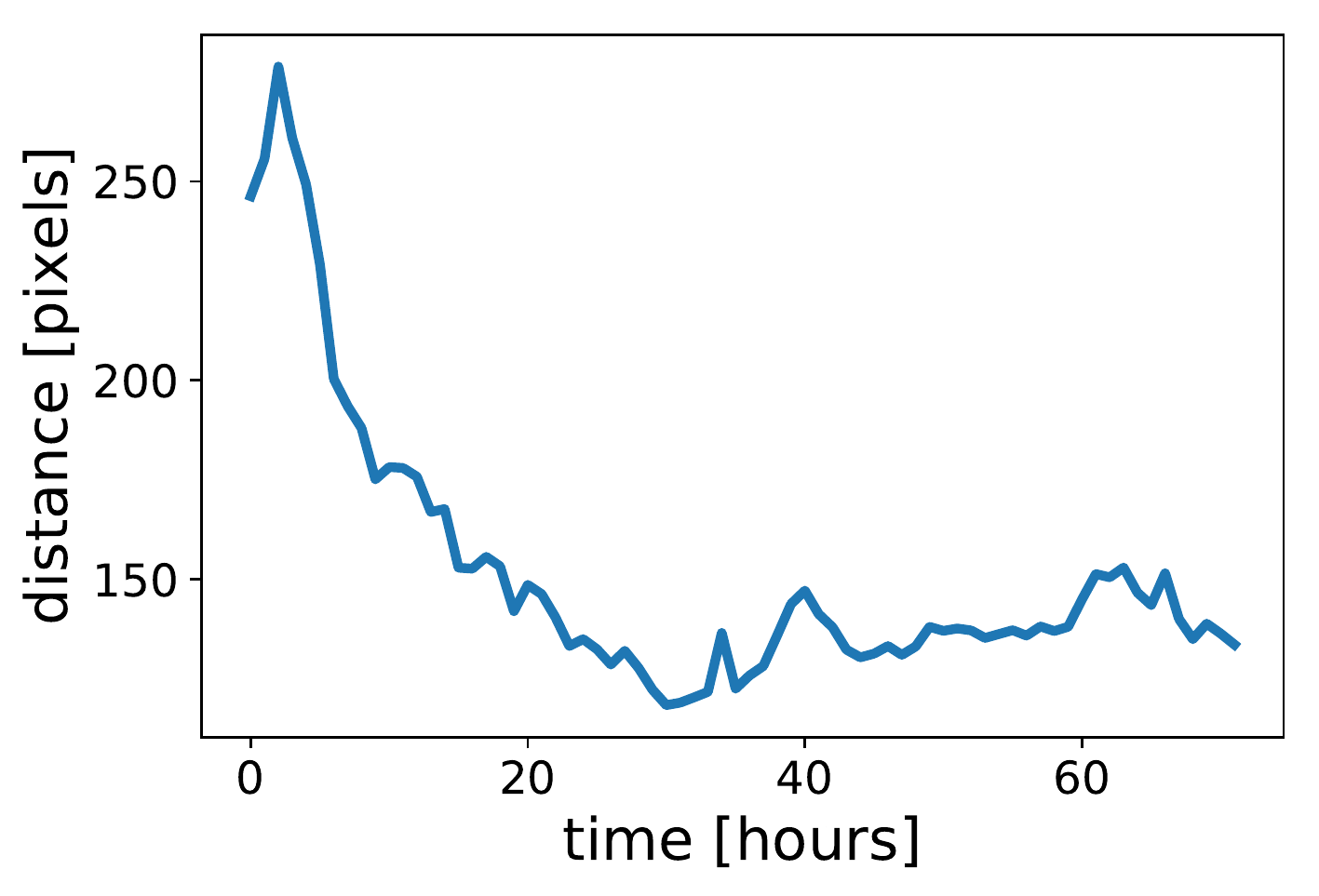}
\end{minipage}
\caption{Percentage of unused space in the canvas, over time (left). Average distance between two subsequent clicks from the same user, over time (right).}
\label{fig:distance_space}
\end{figure}

We pay special attention to the system activity level over time, from its initial chaotic state to the first signs of its convergence. The evolution of the activity level of the entire system is observable through the variations of the click rate. We first notice the level of user activity being clearly influenced by the circadian cycle of American users. Note that, according to Alexa \footnote{https://www.alexa.com/siteinfo/reddit.com}, United States alone represent 57\% of the traffic on Reddit. Beyond the temporal variations of the overall activity within the canvas, the relationship between the number of active users and the number of clicks is of our interest, as it is subject to significant variations that characterize the average activity of single users (see figure~\ref{fig:distribs} right). We observe a rapidly increasing number of click-per-user after the 24 hour. A first potential cause is the coverage of the event (both in social and mainstream media). Moreover, we observe the term ``Reddit~place'' having growth by a factor of $1.8$ from the first to the second day, according to Google Trends. Second, we compare this sudden raise of the activity level with variations of the userbase. At peak value, the number of concurrent unique users per hour was subject to an increase of 27\% from the first to the second day. 

Starting from an empty canvas, users have been continuously filling the space. After the first 24 hours, 90\% of the pixels have already been clicked on at least once (see figure~\ref{fig:distance_space} left). With less blank space at their disposal, users were forced to overwrite existing structure. We observe people having focusing their actions on a similar distribution: the distance between two subsequent clicks performed by the same user has, on average, largely decreased during the first 24 hours (see figure~\ref{fig:distance_space} right).

\begin{figure}[H]
\begin{minipage}[t]{0.48\linewidth}
    \includegraphics[width=\linewidth]{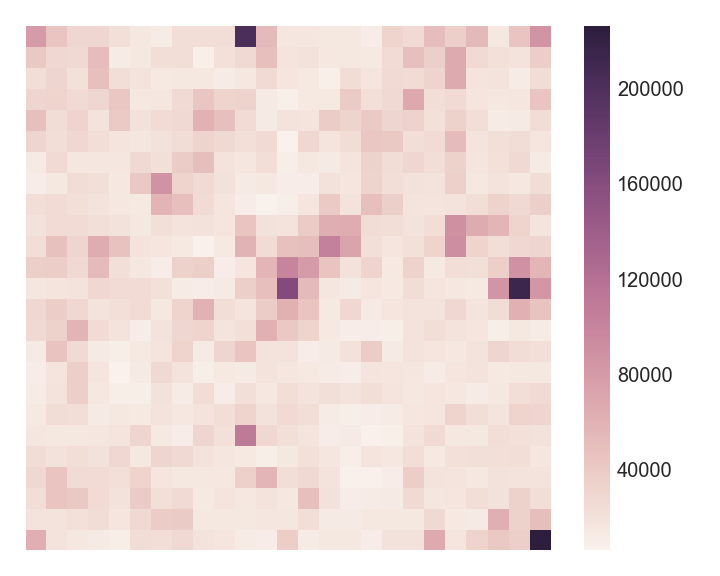}
\end{minipage}%
\hfill%
\begin{minipage}[t]{0.48\linewidth}
    \includegraphics[width=\linewidth]{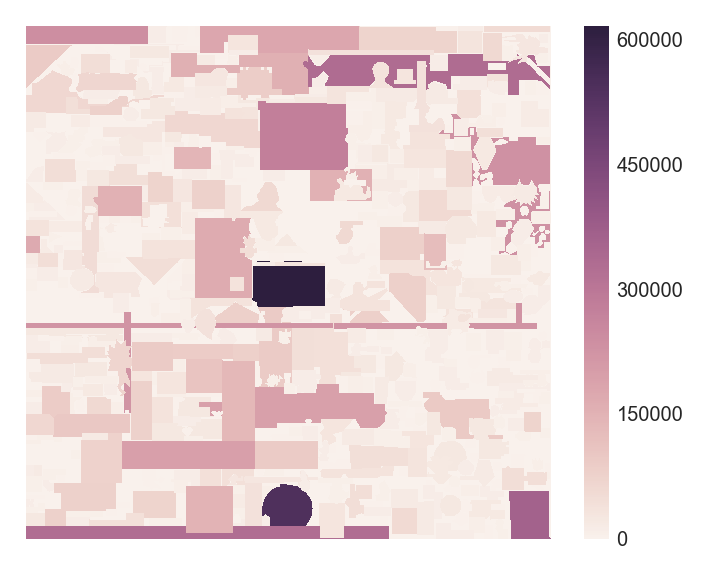}
\end{minipage}
\caption{Activity heatmap on the canvas. On the \textbf{left}, the canvas is partitioned in uniformly-sized tiles. On the \textbf{right}, the canvas regions reflect the artwork shapes, as described in Section~\ref{sec:data}. Both heatmaps use a comparable number of regions.}
\label{fig:heatmaps}
\end{figure}

We observe the activity level with different partitioning schemes of the canvas. Without a clear segmentation following the shape of the artworks, though, only limited observation can be made on the activity taking place on the canvas, i.e., we mainly observe spikes of localized activity (see Figure~\ref{fig:heatmaps} left). However, when considering the artwork shapes, the activity from a group dynamics perspective is revealed (see Figure~\ref{fig:heatmaps} right), therefore motivating the development of an effective segmentation strategy.

\section{Model}
\label{sec:model}

\begin{table}
\centering
\begin{tabular}{lp{0.7\linewidth}}
\toprule
Notation 			& Description                 							\\ \midrule
$U$       			& User set                           					\\
$u_i$     			& User $u_i \in U$ 									\\
$a_k$     			& Action 												\\
$K$       			& Number of latent factors 								\\
$\hat{x}_{u_i a_j}$ & Score of user $u_i$ for action $a_j$ 	\\
$D$  				& Evaluation set	\\
$A_{a_i}$  			& Set of considered users for action $a_i$	\\
$p_{u_i}$  			& Embedding vector for user $u_i$	\\
$q_{a_i}$  			& Joint embedding vector for action $a_i$	\\
\bottomrule
\end{tabular}
\caption{Notation \label{notation-table}}
\end{table}

In this section, we introduce a predictive method that models collaboration between users in order to predict future user actions. In this regard, we train a model to evaluate the likelihood of a user $u_i$ to perform a particular action at a given moment in time. We use the term action as a shorthand, with a slight abuse of notation, to denote the click of a user $u_i$ on a pixel at coordinates $(x,y)$ at time $t$. 

To train the model according to the decision structure of user $u_i$, we train the model to discriminate an action $a_i$ performed by user $u_i$ from a randomly sampled action $a_k$ performed by another user $u_k \neq u_i$. We define it as a probability 

\begin{equation}
\Pr(a_i>_{u_i}a_j|\Theta),
\end{equation}

where $\Theta$ represents the set of parameters of an arbitrary predictor and $>_{u_i}$ represents the preference scheme of user $u_i$. Specifically, we use the notation $a_i>_{u_i}a_j$ to indicate user $u_i$ preferring action $a_i$ over action $a_j$. A predictor that would perfectly model the latent preference structure $>_{u_i}$ of user $u_i$ would thus predict a probability of 1 for ${\Pr(a_i>_{u_i}a_j|\Theta)}$ and a probability of 0 for ${\Pr(a_i<_{u_i}a_j|\Theta)}$. Defining $\hat{x}_{u_i,a_i}$ as the predicted score for user $u_i$ and action $a_i$, the same ideal predictor would systematically predict a higher score for $\hat{x}_{u_i,a_i}$ than for $\hat{x}_{u_i,a_j}$, thus making the quantity $\hat{x}_{u_i,a_i,a_j} := \hat{x}_{u_i,a_i} - \hat{x}_{u_i,a_j}$ consistently and strictly positive. Therefore, training the model to discriminate between the two actions can be achieved by maximizing the difference between their predicted scores. In order to make the operation differentiable, we follow the procedure introduced by Rendle et al. \cite{Rendle2009BPRBP} by maximizing the quantity $\ln\sigma(\hat{x}_{u_i,a_i,a_j})$. In particular, we maximize the following criterion

\begin{equation}
\text{BPR-OPT} := \sum_{(u_i, a_i, a_j)\in D} \ln\sigma(\hat{x}_{u_i,a_i,a_j}) - \lambda_{\Theta} ||\Theta||^2
\end{equation}

where $\lambda_{\Theta}$ is a parameter controlling the strength of the regularization. In our case, we opted for $\ell_2$-regularization.

So far, we described our optimization criterion without specifying the underlying predictive model. Our choice of predictor is driven by its capability to model users-users relationships. We opt for an embedding method, since we hypothesize less independent behaviors than individuals in the system. Embedding methods are especially adapted to produce \textit{personalized} predictions (e.g. collaborative filtering applications), by making the assumption that the behavior from an individual can be predicted by collecting data from many users and by projecting each of them in a common latent space. We therefore represent every user in the system by a latent representation: a real-valued vector $p_{u_i}$ of size $K$ where $K$ is the chosen dimensionality of the latent space. We define a notion of distance between any pair of users in the considered population, where the distance metric represents the strength of collaboration between users. If two users are actively collaborating, the response produced by the combination of their respective vectors (typically by using dot product) should be high.

\begin{figure}
	\centering
    \includegraphics[width=0.7\linewidth]{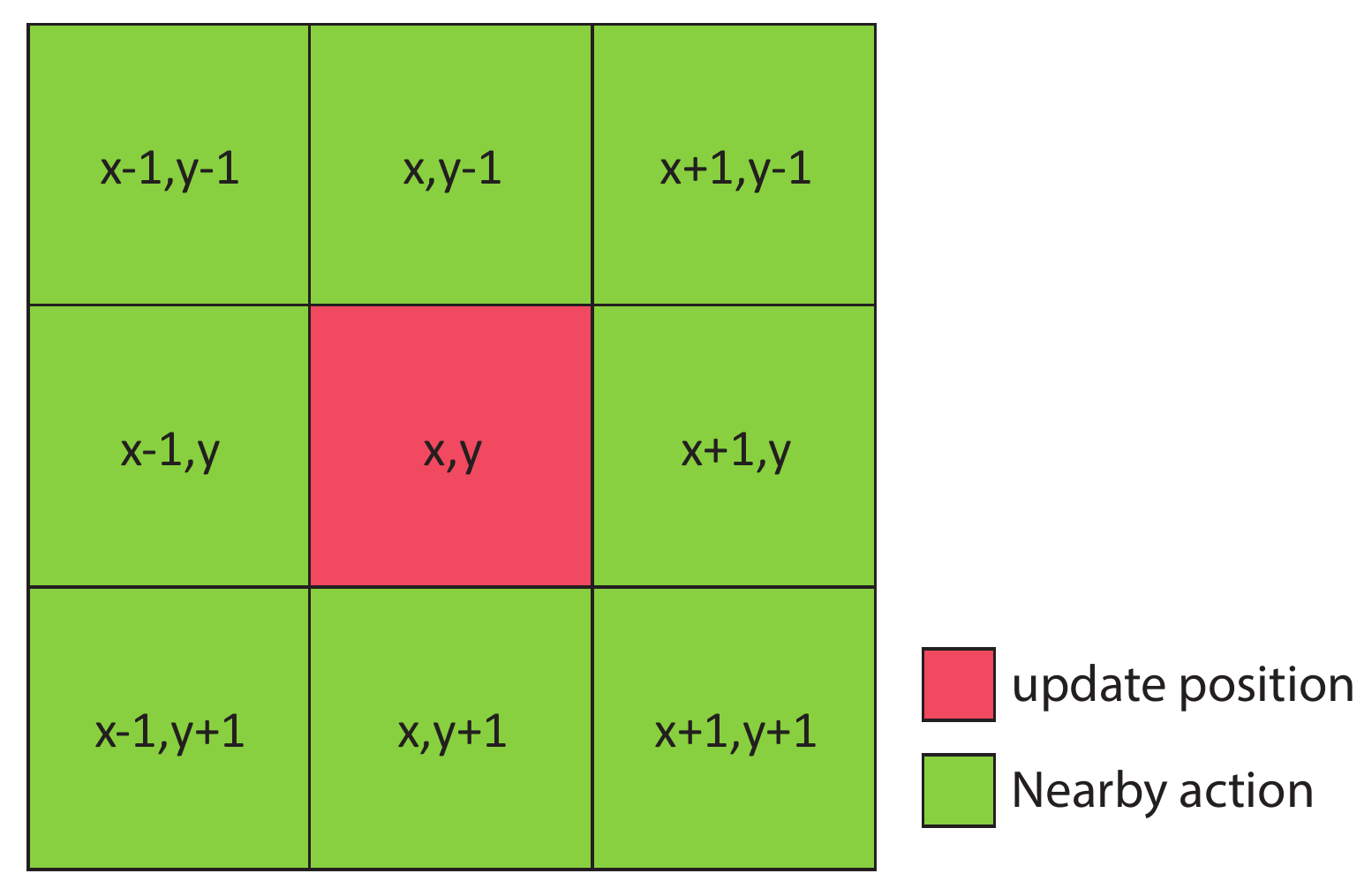}
	\caption{Illustration of the input features. For an action performed on pixel (x,y), we consider the users having updated the 8 adjacent pixels at last.}
    \label{fig:illustr}
\end{figure}

As suggested by previous studies (see section~\ref{sec:related}), the choice of input features is determinant to capture complex inter-user dependencies. Assuming users performing actions in a fully-observable environment, we construct our input signal by observing the proximity of their actions. In our scenario, the notion of proximity could be constructed by looking at the tiles surrounding the one on which the considered user $u_i$ clicked (see figure~\ref{fig:illustr}). Assuming every user being represented by a latent representation vector $p_{u_i}$, we define $q_{a_i}$ to be a combination of the users' latent features vectors that updated the eight adjacent cells at last, before action $a_i$. We then specifically train our model to produce a high predicted score $\hat{x}_{u_i,a_i}$ of user $u_i$ performing action $a_i$. The score $\hat{x}_{u_i,a_i}$ is computed as follows

\begin{equation}
\hat{x}_{u_i,a_i} = p_{u_i}^T \cdot q_{a_i}
\end{equation}

where $q_{a_i}$ is a combination of users' embeddings. This combination could be computed in many ways, using a differentiable operation. We combined the embedding vectors using a simple sum over users' latent representations, defined as follows

\begin{equation}
q_{a_i} = \sum_{k \in A_{a_i}} p_{u_k}
\end{equation}

where $A_{a_i}$ is the list of users having updated the adjacent tiles at last, before action $a_i$. During our experiments, we observed the normalization of each vector $p_{u_k}$ to benefit from normalization (in the sense of $\ell_2$ normalization). 

\subsection{Optimization}
\label{sec:opti}
As we aim to discriminate positive from negative examples, we devise our optimization procedure as a ranking problem. In particular, we train the predictor to discriminate an observed interaction from a randomly sampled negative (an action performed by another user). The BPR optimization scheme can be optimized using the following update step

\begin{equation}
	\label{eq:bpr}
	\theta \leftarrow \theta + \alpha \cdot (\sigma(-\hat{x}_{u_i,a_i,a_j}) \
    \frac{\partial\hat{x}_{u_i,a_i,a_j}}{\partial \theta} + \lambda_{\theta}\Omega'(\theta)),
\end{equation}

where $\hat{x}_{u_i,a_i,a_j} = \hat{x}_{u_i a_i} - \hat{x}_{u_i a_j}$, and $\theta$ represents the set of parameters to be learned. $\Omega(\theta)$ denotes a regularizer. We opted for a $\ell_2$ regularization $\Omega(\theta) = \| \Theta \|_2^2$.

\begin{figure*}
\begin{minipage}[t]{0.48\linewidth}
    \includegraphics[width=\linewidth]{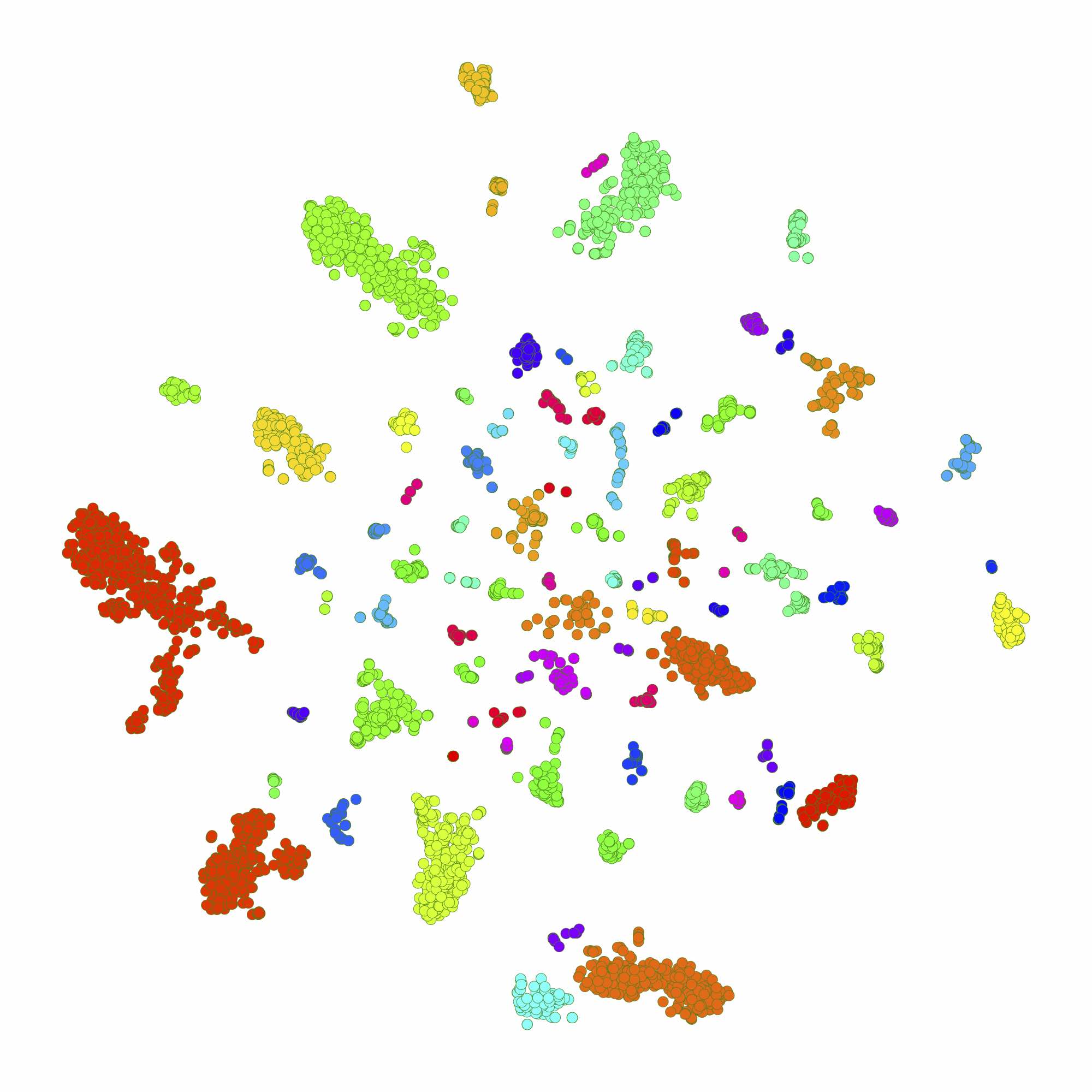}
\end{minipage}%
\hfill%
\begin{minipage}[t]{0.48\linewidth}
    \includegraphics[width=\linewidth]{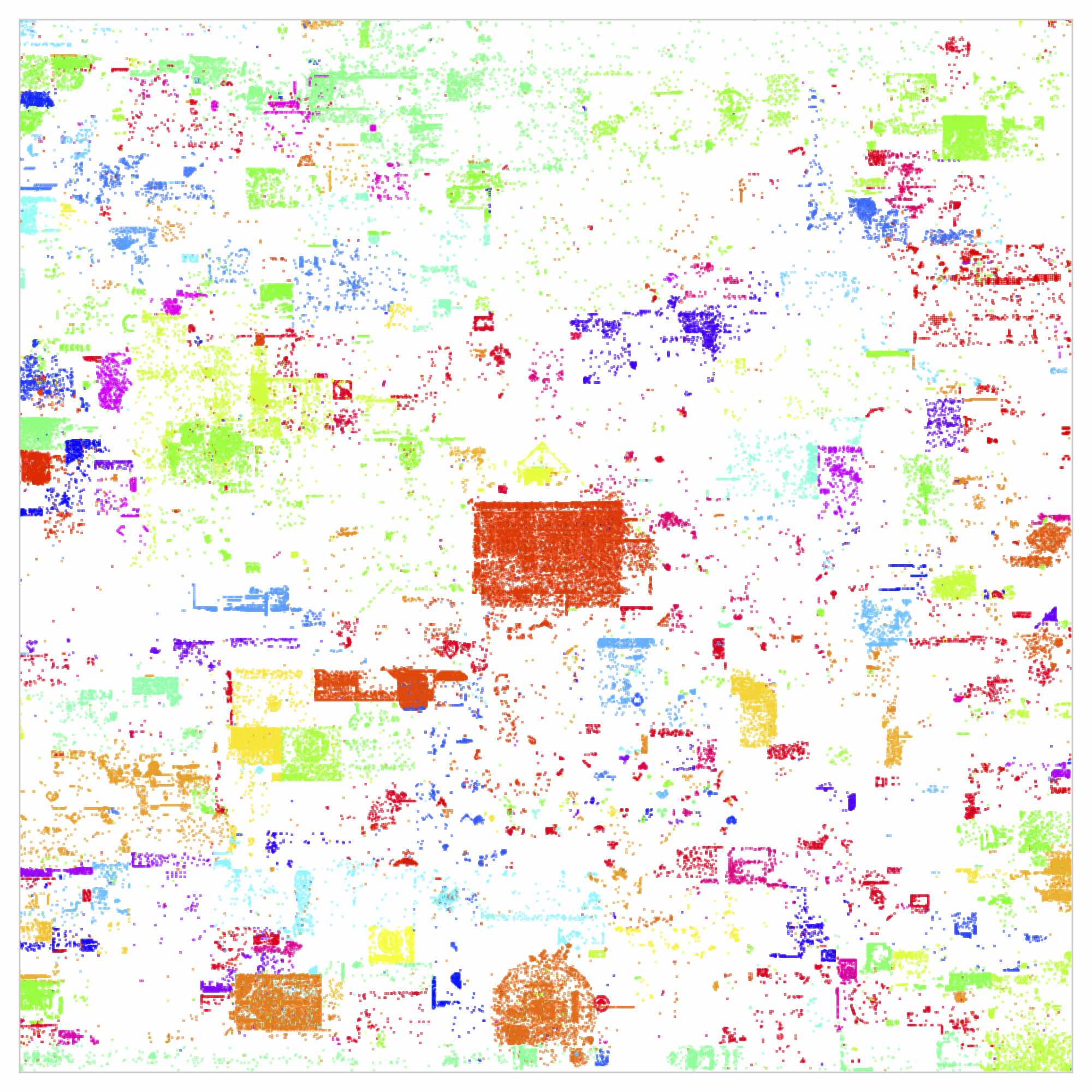}
\end{minipage}
\caption{Collaboration patterns on Reddit r/place. Groups of users have been identified by exploiting the computed latent representation (\textbf{left}), and their activities have been localized over the canvas (\textbf{right}). Best seen in colors. \emph{Left:} t-SNE \cite{Maaten2008VisualizingDU} projection of user embeddings. For visualization purposes, user groups have been colored using the resulting clustering from DBSCAN \cite{Ester1996ADA}. \emph{Right:} Traces of activity performed by selected users. Colors correspond to the left figure. Results are computed for the last 1M interactions (around 3 hours of activity).}
\label{fig:tsne}
\end{figure*}

\subsection{Experimental Setting}
\label{sec:experimental}
We adopt a \textit{leave-one-out} methodology to assess the accuracy of the model, thus making every user having the same weight in the evaluation. Specifically, we constitute our evaluation set by sampling one action for each user. As an experimental setting, we discard users having performed less than 10 interactions from our dataset and filter the first quarter of interactions. We apply this filtering to avoid a cold-start data regime, a scenario that is outside of the scope of this work.

\xhdr{Reproducibility:} We ran our experiment on a single computer, running a 3.2 GHz Intel Core i7 CPU, using PyTorch version 0.2.0.4 \footnote{http://pytorch.org/}. We run the optimization on GPU NVIDIA GTX 670. We trained our model with the following parameters: $\alpha = 0.04$, $\lambda_{\theta} = 0.01$, $K = 120$. All code will be made available at publication time~\footnote{https://github.com/JRappaz/placemf}.

\subsection{Baselines:}
We describe the baselines to which our approach is compared to. Those methods are split in two different categories: methods that model the interaction of users with their environment and methods that model users interrelationships.

\begin{description}
\item \textbf{Median}: As discussed in section~\ref{sec:analysis}, users have rapidly focused their actions on specific areas of the canvas. We therefore compare our results to a simple baseline that compute the likelihood of interaction as a linear function of the euclidean distance between two points $p_1$ and $p_2$, where $p_1$ is the position of the considered click and $p_2$ is the median position of the user activity in the canvas (computed from the training set). We also tried with the average position but exclude the results as they were consistently lower.

\item \textbf{MF}: Matrix Factorization (MF) methods typically model the preferences from users interacting with a large number of items. We trained the MF baseline to model the interactions between users and pixels, in a setting similar to a preference problem. We used the Spark version 2.2 ALS implementation, a scalable model of Collaborative Filtering method. \footnote{https://spark.apache.org/}.
\item  \textbf{Count}: We compare our method to a simple count of user interactions. We count the number of adjacent clicks between users and rank the available actions based on the location having the largest sum of users' interactions.
\item  \textbf{Community Detection}: We use Infomap\footnote{http://www.mapequation.org/code.html}, a scalable, state-of-the-art community detection algorithm optimizing an information-theoretic criterion. We model collaboration between user as a weighted undirected graph. Edges weights are computed from the number of past adjacent clicks between two users. The algorithm provides a single assignment for each user to a cluster. The predicted score for an action and a user $u_i$ is then computed by counting the number of users having updated the adjacent tiles at last and being part of the same community than $u_i$.

\end{description}

\subsection{Evaluation} \label{subsec:eval}
To evaluate our approach, we select the widely used metric \textit{Area Under the Curve} (AUC)~\cite{shani2011evaluating} as our measure of performance:
\begin{equation} 
\label{eq1}
\text{AUC} = \frac{1}{\left\vert{D}\right\vert} \sum_{\!\!\!\!\!\!\!\!\!\!\!\!\mathrlap{(u_i, a_i, a_j)\in D}} H( \hat{x}_{u_i a_i} - \hat{x}_{u_i a_j})
		   = \frac{1}{\left\vert{D}\right\vert} \sum_{\!\!\!\!\!\!\!\!\!\!\!\!\mathrlap{(u_i, a_i, a_j)\in D}} H( \hat{x}_{u_i, a_i, a_j}), 
\end{equation}

where $H(\cdot)$ is the Heaviside step function (equal to 1 for a positive input, zero otherwise) and $D$ is our evaluation set composed of one triplet $(u_i, a_i, a_j)$ per source where $u_i$ is a user, $a_i$ is a randomly sampled action that has been performed by user $u_i$ and $a_j$ is a randomly sampled action from another user $u_j \neq u_i$. This metric assesses the ability of the predictor to correctly rank a positive interaction withheld during training against a random negative example. Negative examples are sampled from the training set during training and from the testing set during testing. An ideal predictor would obtain a score of $AUC=1$, while a random selection would output a score around $AUC=0.5$.

\section{Results}
\label{sec:results}

\begin{table}[]
\centering

\begin{tabular}{lll}
\hline
Method      &       & AUC                 \\ \hline
Environment & Median& 0.8413 $\pm$ 0.0006 \\
            & MF    & 0.7921 $\pm$ 0.0006 \\ \hline
Social      & Count & 0.8383 $\pm$ 0.0003 \\
            & CD    & 0.8382 $\pm$ 0.0006 \\
            & Ours  & \textbf{0.8792} $\pm$ 0.0019 \\ \hline
\end{tabular}
\caption{Results: scores are reported with the Area Under the Curve (AUC) metric (CI=0.95).}
\label{tab:results}
\end{table}

In this section, we discuss the results summarized in table~\ref{tab:results}. We divide the aforementioned methods in two categories: methods that capture the relation between users and their environment, and methods that capture users' inter-relationships.

\begin{figure*}[t]
    \includegraphics[width=\linewidth]{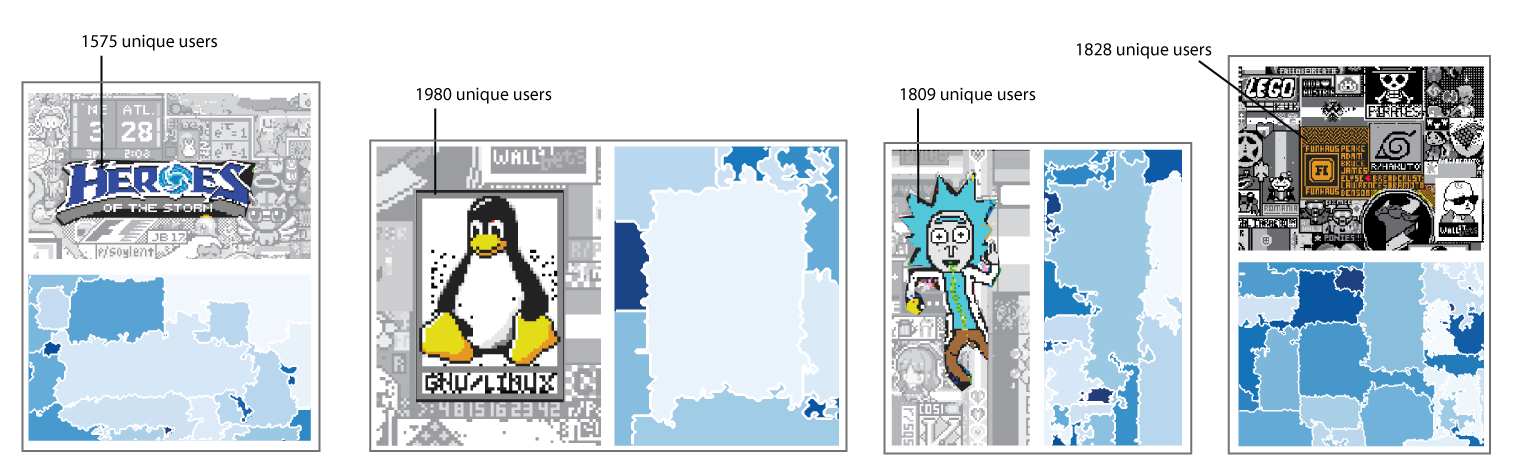}
	\caption{Segmentation of the artworks using the proposed method which leverages exclusively social signals. We report the number of users that contributed to the final version of the highlighted artworks. Even with artworks produced by a large user base, our segmentation method can correctly identify the boundaries, as shown in the above examples.}
	\label{fig:segmentation}
\end{figure*}

\xhdr{User-environment relationship:} The virtual environment is constituted by a set of clickable tiles. Moreover, this environment is structured, since pixels have clearly defined positions. MF-based methods are unaware of the structure of the environment and did not model any social aspect of the event. We therefore observe them to exhibit a lower score than the other methods. The score obtained from a median-based method is, however, performing relatively well, despite its relative simplicity. This suggest the locality of user actions being a critical aspect in the design of a method to predict collaborations.

\xhdr{User-user relationship:} We observe the Community Detection method being on par with a method based on raw interactions count. We therefore suggest that a strict segmentation of the users does not result on a gain in performance. Since those methods were able to capture relatively good social proximity between users, they were not optimized (and not parametrized) to directly model user interactions. From table~\ref{tab:results}, we can observe the performance gain obtained from our model, being optimized to model user interactions data. Specifically, our methods outperformed the best performing baseline with a margin of 4.5\% of relative improvement. We therefore conclude that, from the considered models, the parametrization of user interrelationships is the most predictive method of user actions in a sandbox environment. 

Systems are generally considered as complex if the sum of the individual behaviors of its subparts cannot explain the overall behavior of the ensemble. This consideration encourages the modeling of the inter-relationships between subparts, instead of modeling them as independent behaviors. Our results reinforce this perspective, as we report this modeling approach to be more predictive of users' decisions. Moreover, our approach has the advantage to be interpretable, as described in the following section.

\section{Segmentation}
\label{sec:segmentation}
In this section, we exploit the learned representation of our model to identify user groups and segment their respective artworks. 

First we show that groups of users can be identified from the latent representation obtained through the proposed method. As described in section~\ref{sec:model}, each user is represented by a single low-dimensional embedding vector $p_{u_i}$. The total set of vectors represents a distributed representation of the collaboration strength between any two considered users (all the vectors share the same latent space but the triangle inequality is not necessarily respected). This representation can be further reduced in dimensionality in order to be visualized. We observe the resulting representation to exhibit sufficient cohesiveness to be easily labeled by an unsupervised clustering approach. The traces left in the canvas by the different groups of users is shown in figure~\ref{fig:tsne}.

\begin{figure}[htb]
	\centering
    \includegraphics[width=0.7\linewidth]{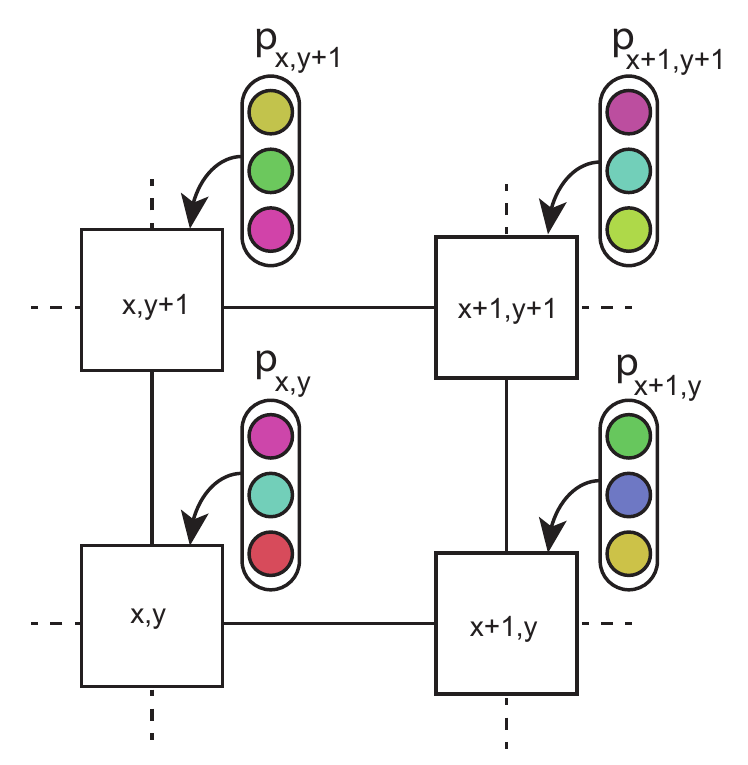}
	\caption{Illustration of the agglomerative clustering procedure setting. Each pixel has a latent features vector attached to it. Black lines represent the connectivity constraint applied to the procedure: only adjacent pixels can be merged.}
    \label{fig:linkage}
\end{figure}

In a second step, we propose a method to segment the product of user collaboration, i.e. to attribute every pixel to a single partition of the final image. To this end, we propose a simple method that detects abrupt variations of the social activity on the canvas. We first assume that users leave a social \textit{signature} in the canvas, and suggest that abrupt variations of this signature could reveal the artworks edges. To model it, we attach to each pixel a fingerprint of the social activity that took place on it. We opt for the latent representation of the user that updated the pixel at last, i.e., the user whose action colored a pixel of the final canvas. We then segment the canvas by using an agglomerative clustering procedure. At start, the algorithm attributes each pixel to its own cluster. Then, the procedure recursively merges clusters by minimizing the sum of squared differences within all clusters. Note that the algorithm is constrained by the connectivity of the grid as only adjacent pixels could be merged (see figure~\ref{fig:linkage}). The procedure terminates when the target number of clusters $C$ is obtained. To find the optimal value of the $C$ parameter, we rely on the crowd-sourced segmentation of user artwork that we treat as ground truth. We search for the optimal number of clusters, restricting our evaluation to the portions of the space being annotated. We use Adjusted Rand Index (ARI), as a metric comparing the results of the clustering procedure with the annotations. We found the best value to be $C=\boldsymbol{840}$, a value close to the 830 artworks identified in human-curated atlas. Examples of the resulting clustering are shown in figure~\ref{fig:segmentation}.

\section{Discussion}

The method introduced in Section~\ref{sec:model} exhibits a significant improvement over the baselines in a purely predictive task. We suggest that this result is due to its capability of modeling user-user relationships, while still capturing local consistency of the environment. Indeed, our approach directly models a 8-to-1 relationship between the considered user and the users having updated the adjacent tiles. Moreover, thanks to the common latent space in which all the users are projected, our method models transitive relations between participants. As an example, if user A collaborated with user B and user B with user C, a potential collaboration between A and C should still exhibit a high predicted score since they would be contained in the same local manifold of the embedding space. This is especially important in the case of large artworks, to which many users contributed.

We propose a qualitative interpretation of the learned parameters in figure~\ref{fig:segmentation}, where we show the traces left by participants on the canvas. We observe the method to capture a relative proximity of actions performed by users belonging to the same group. This result is coherent with the observation of user actions being relatively localized (see section~\ref{sec:analysis}).

The proposed method captures the proximity of users' actions on the canvas. One might question the ability of such approach to establish a clear segmentation of the artworks from proximity signals. For example, two different user populations, working on two different adjacent artworks, are difficult to be distinguished, as the actions of the two groups would appear in close vicinity. However, we give evidences that a latent representation, computed from a large set of actions, is sufficiently expressive to recover the artworks boundaries.

\section{Conclusion and Future Work}
We introduce a generic method to infer collaboration patterns in environments where only user interactions are observable. We show, through experiments, that the local proximity of users' actions represents a sufficiently expressive signal for the study of collaboration. Indeed, we report it to be more predictive than the modeling of the interactions between users and their environment. This finding reinforces previous results in the domain, that suggest the study of emergent phenomenons requiring the modeling of interrelationships between the parts of a system, rather than modeling their individual behaviors.

Being able to capture rich social signals, such as collaboration patterns, represents a unique opportunity to study complex social phenomenons. With virtual environments being increasingly used as a proxy to study specific social-psychology aspects, our method paves the way for the analysis of user behavior in contexts in which only user actions are observable, and the collaboration patterns are emergent rather than predefined.

Our method finds immediate applications in the analysis of large-scale collaborative efforts, such as Wikipedia or Github, in which users don't always have explicit or observable links between them. In such cases, the proper segmentation of user contributions could lead to a more fine-grained quantitative analysis of the various portions of the contribution. In concrete terms, the method would allow assessing the quality of the contribution produced by a subgroup of the population instead of measuring the quality of the article/repository as a whole. Beyond the analysis of collaboration, such methods could also be used to produce recommendations of partnership as a way to engage participants through direct collaboration recommendation. Our method can finally find applications in malicious collaboration. In such context, the result of a collaboration could be isolated using similar segmentation techniques to limit the impact of adversarial activities on large projects.

We foresee several directions as future work.
\begin{itemize}
\item Given the generic nature of our method, we want to test it on other collaborative platforms (e.g., Wikipedia, Github, etc.) as they represent a natural extension of our current effort on Reddit r/place. Moreover, we believe that other types of virtual environments, such as multi-player video-games, would represent an interesting testbed for the proposed method. 
\item We want to tackle the problem of learning in a low data regime (i.e., cold-start scenario). We plan to leverage side-information, inherent to both users and their environment, as it represents a promising resource to generate prediction for newly introduced users. 
\item We want to make our model temporally-aware, as further insights can be gathered by analyzing the temporal dynamics of the user interactions.
\item We plan to study the dual problem, that is to infer antisocial behaviors from localized actions, and to propose a method to distinguish them from collaboration patterns.
\end{itemize}

\clearpage
\bibliographystyle{aaai}
\bibliography{references} 

\begin{thebibliography}{}

\bibitem[\protect\citeauthoryear{Bar-Yam}{2002}]{bar2002general}
Bar-Yam, Y.
\newblock 2002.
\newblock General features of complex systems.
\newblock {\em Encyclopedia of Life Support Systems (EOLSS), UNESCO, EOLSS
  Publishers, Oxford, UK}.

\bibitem[\protect\citeauthoryear{Bell and Koren}{2007}]{Bell2007TheBS}
Bell, R.~M., and Koren, Y.
\newblock 2007.
\newblock The bellkor solution to the netflix prize.

\bibitem[\protect\citeauthoryear{Cole and Griffiths}{2007}]{cole2007social}
Cole, H., and Griffiths, M.~D.
\newblock 2007.
\newblock Social interactions in massively multiplayer online role-playing
  gamers.
\newblock {\em CyberPsychology \& Behavior} 10(4):575--583.

\bibitem[\protect\citeauthoryear{de Montjoye \bgroup et al\mbox.\egroup
  }{2014}]{Montjoye2014TheSO}
de~Montjoye, Y.-A.; Stopczynski, A.; Shmueli, E.; Pentland, A.; and Lehmann, S.
\newblock 2014.
\newblock The strength of the strongest ties in collaborative problem solving.
\newblock In {\em Scientific reports}.

\bibitem[\protect\citeauthoryear{Ester, Kriegel, and J{\"{o}rg Sander and
  Xiaowei Xu}}{}]{Ester1996ADA}
Ester, M.; Kriegel, H.-P.; and J{\"{o}rg Sander and Xiaowei Xu}, y.
\newblock A density-based algorithm for discovering clusters in large spatial
  databases with noise.

\bibitem[\protect\citeauthoryear{Gershenson and
  Fern{\'a}ndez}{2012}]{Gershenson2012ComplexityAI}
Gershenson, C., and Fern{\'a}ndez, N.
\newblock 2012.
\newblock Complexity and information: Measuring emergence, self-organization,
  and homeostasis at multiple scales.
\newblock {\em Complexity} 18:29--44.

\bibitem[\protect\citeauthoryear{Grover and
  Leskovec}{2016}]{grover2016node2vec}
Grover, A., and Leskovec, J.
\newblock 2016.
\newblock node2vec: Scalable feature learning for networks.
\newblock In {\em Proceedings of the 22nd ACM SIGKDD international conference
  on Knowledge discovery and data mining},  855--864.
\newblock ACM.

\bibitem[\protect\citeauthoryear{Kearns}{2012}]{kearns2012experiments}
Kearns, M.
\newblock 2012.
\newblock Experiments in social computation.
\newblock {\em Communications of the ACM} 55(10):56--67.

\bibitem[\protect\citeauthoryear{Kittur and Kraut}{2008}]{kittur2008harnessing}
Kittur, A., and Kraut, R.~E.
\newblock 2008.
\newblock Harnessing the wisdom of crowds in wikipedia: quality through
  coordination.
\newblock In {\em Proceedings of the 2008 ACM conference on Computer supported
  cooperative work},  37--46.
\newblock ACM.

\bibitem[\protect\citeauthoryear{Kozlov and Johansen}{2010}]{Kozlov2010RealBI}
Kozlov, M.~D., and Johansen, M.~K.
\newblock 2010.
\newblock Real behavior in virtual environments: Psychology experiments in a
  simple virtual-reality paradigm using video games.
\newblock {\em Cyberpsychology, behavior and social networking} 13 6:711--4.

\bibitem[\protect\citeauthoryear{KubÃ­}{2003}]{doi:10.1162/106454603321489518}
KubÃ­, A.
\newblock 2003.
\newblock Toward a formalization of emergence.
\newblock {\em Artificial Life} 9(1):41--65.

\bibitem[\protect\citeauthoryear{Majchrzak, Malhotra, and
  John}{2005}]{Majchrzak2005PerceivedIC}
Majchrzak, A.; Malhotra, A.; and John, R.
\newblock 2005.
\newblock Perceived individual collaboration know-how development through
  information technology - enabled contextualization: Evidence from distributed
  teams.
\newblock {\em Information Systems Research} 16:9--27.

\bibitem[\protect\citeauthoryear{Mason and
  Watts}{2012}]{mason2012collaborative}
Mason, W., and Watts, D.~J.
\newblock 2012.
\newblock Collaborative learning in networks.
\newblock {\em Proceedings of the National Academy of Sciences}
  109(3):764--769.

\bibitem[\protect\citeauthoryear{Mataric}{1993}]{mataric1993designing}
Mataric, M.~J.
\newblock 1993.
\newblock Designing emergent behaviors: From local interactions to collective
  intelligence.
\newblock In {\em Proceedings of the Second International Conference on
  Simulation of Adaptive Behavior},  432--441.

\bibitem[\protect\citeauthoryear{Mersch, Crespi, and Keller}{2013}]{Mersch1090}
Mersch, D.~P.; Crespi, A.; and Keller, L.
\newblock 2013.
\newblock Tracking individuals shows spatial fidelity is a key regulator of ant
  social organization.
\newblock {\em Science} 340(6136):1090--1093.

\bibitem[\protect\citeauthoryear{Newman}{2001}]{Newman2001ScientificCN}
Newman, M. E.~J.
\newblock 2001.
\newblock Scientific collaboration networks. ii. shortest paths, weighted
  networks, and centrality.
\newblock {\em Physical review. E, Statistical, nonlinear, and soft matter
  physics}.

\bibitem[\protect\citeauthoryear{Pan \bgroup et al\mbox.\egroup
  }{2008}]{Pan2008OneClassCF}
Pan, R.; Zhou, Y.; Cao, B.; Liu, N.~N.; Lukose, R.~M.; Scholz, M.; and Yang, Q.
\newblock 2008.
\newblock One-class collaborative filtering.
\newblock {\em 2008 Eighth IEEE International Conference on Data Mining}
  502--511.

\bibitem[\protect\citeauthoryear{Ramasco, Dorogovtsev, and
  Pastor-Satorras}{2004}]{ramasco2004self}
Ramasco, J.~J.; Dorogovtsev, S.~N.; and Pastor-Satorras, R.
\newblock 2004.
\newblock Self-organization of collaboration networks.
\newblock {\em Physical review E} 70(3):036106.

\bibitem[\protect\citeauthoryear{Rendle \bgroup et al\mbox.\egroup
  }{2009}]{Rendle2009BPRBP}
Rendle, S.; Freudenthaler, C.; Gantner, Z.; and Schmidt-Thieme, L.
\newblock 2009.
\newblock Bpr: Bayesian personalized ranking from implicit feedback.
\newblock In {\em UAI}.

\bibitem[\protect\citeauthoryear{Rosvall and
  Bergstrom}{2008}]{Rosvall2008MapsOR}
Rosvall, M., and Bergstrom, C.~T.
\newblock 2008.
\newblock Maps of random walks on complex networks reveal community structure.
\newblock {\em Proceedings of the National Academy of Sciences of the United
  States of America} 105 4:1118--23.

\bibitem[\protect\citeauthoryear{Sekara and Lehmann}{2014}]{Sekara2014TheSO}
Sekara, V., and Lehmann, S.
\newblock 2014.
\newblock The strength of friendship ties in proximity sensor data.
\newblock In {\em PloS one}.

\bibitem[\protect\citeauthoryear{Shani and
  Gunawardana}{2011}]{shani2011evaluating}
Shani, G., and Gunawardana, A.
\newblock 2011.
\newblock Evaluating recommendation systems.
\newblock In {\em Recommender systems handbook}. Springer.
\newblock  257--297.

\bibitem[\protect\citeauthoryear{Suri and Watts}{2011}]{suri2011cooperation}
Suri, S., and Watts, D.~J.
\newblock 2011.
\newblock Cooperation and contagion in web-based, networked public goods
  experiments.
\newblock {\em PloS one} 6(3):e16836.

\bibitem[\protect\citeauthoryear{Tang \bgroup et al\mbox.\egroup
  }{2012}]{Tang2012CrossdomainCR}
Tang, J.; Wu, S.; Sun, J.; and Su, H.
\newblock 2012.
\newblock Cross-domain collaboration recommendation.
\newblock In {\em KDD}.

\bibitem[\protect\citeauthoryear{van~der Maaten, Hinton, and
  Bengio}{2008}]{Maaten2008VisualizingDU}
van~der Maaten, L.; Hinton, G.; and Bengio, Y.
\newblock 2008.
\newblock Visualizing data using t-sne.

\bibitem[\protect\citeauthoryear{Yang and
  Leskovec}{2013}]{Yang2013OverlappingCD}
Yang, J., and Leskovec, J.
\newblock 2013.
\newblock Overlapping community detection at scale: a nonnegative matrix
  factorization approach.
\newblock In {\em WSDM}.

\bibitem[\protect\citeauthoryear{Zhang and
  Yeung}{2012}]{Zhang2012OverlappingCD}
Zhang, Y., and Yeung, D.-Y.
\newblock 2012.
\newblock Overlapping community detection via bounded nonnegative matrix
  tri-factorization.
\newblock In {\em KDD}.

\end{thebibliography}

\end{document}